\documentclass[conference]{IEEEtran}
\IEEEoverridecommandlockouts
\usepackage{cite}
\usepackage{amsmath,amssymb,amsfonts}
\usepackage{algorithmic}
\usepackage{graphicx}
\usepackage{textcomp}
\usepackage{xcolor}
\usepackage{booktabs}
\usepackage{array}
\usepackage{makecell}
\usepackage{adjustbox}
\def\BibTeX{{\rm B\kern-.05em{\sc i\kern-.025em b}\kern-.08em
    T\kern-.1667em\lower.7ex\hbox{E}\kern-.125emX}}
\begin{document}

\title{A Non-Invasive Interpretable NAFLD Diagnostic Method Combining TCM Tongue Features
}

\author{\IEEEauthorblockN{Shan Cao}
\IEEEauthorblockA{\textit{Xiamen University}\\
Xiamen, China \\
30920201153954@stu.xmu.edu.cn}
\and
\IEEEauthorblockN{Qunsheng Ruan}
\IEEEauthorblockA{\textit{Ganzhou Teachers College}\\
Ganzhou, China \\
ruanqunsheng1979@163.com}
\and
\IEEEauthorblockN{Qingfeng Wu*}
\IEEEauthorblockA{\textit{Xiamen University}\\
Xiamen, China \\
*Correspondence Author: qfwu@xmu.edu.cn}
\and
\IEEEauthorblockN{Weiqiang Lin}
\IEEEauthorblockA{\textit{Xiamen University}\\ 
Xiamen, China \\
wqlin@xmu.edu.cn}
}

\maketitle

\begin{abstract}
Non-alcoholic fatty liver disease (NAFLD) is a clinicopathological syndrome characterized by hepatic steatosis resulting from the exclusion of alcohol and other identifiable liver-damaging factors. It has emerged as a leading cause of chronic liver disease worldwide. Currently, the conventional methods for NAFLD detection are expensive and not suitable for users to perform daily diagnostics. To address this issue, this study proposes a non-invasive and interpretable NAFLD diagnostic method, the required user-provided indicators are only Gender, Age, Height, Weight, Waist Circumference, Hip Circumference, and tongue image. This method involves merging patients' physiological indicators with tongue features, which are then input into a fusion network named SelectorNet. SelectorNet combines attention mechanisms with feature selection mechanisms, enabling it to autonomously learn the ability to select important features. The experimental results show that the proposed method achieves an accuracy of 77.22\% using only non-invasive data, and it also provides compelling interpretability matrices. This study contributes to the early diagnosis of NAFLD and the intelligent advancement of TCM tongue diagnosis. The project mentioned in this paper is currently publicly available\footnote{https://github.com/cshan-github/SelectorNet}.
\end{abstract}


\section{Introduction}

The prevalence of NAFLD has been on the rise in recent years. In the United States (64 million patients) and four major European countries (52 million), the annual healthcare costs attributed to NAFLD are approximately 150 billion dollars\cite{b1}. While many NAFLD patients are asymptomatic, the disease can progress to fibrosis, which is considered a significant contributor to severe liver diseases and mortality in affected individuals \cite{b2,b3,b4}. Timely diagnosis of NAFLD can facilitate essential lifestyle interventions, aiding in preventing disease progression and reversing hepatic lipid accumulation.

Currently, the diagnostic methods for NAFLD include ultrasound, computed tomography (CT) scan, liver biopsy and so on\cite{b5}. While these methods can indeed provide accurate diagnostic results, their cost and limited flexibility significantly impact their feasibility for large-scale and long-term diagnosis. Therefore, developing a low-cost diagnostic method for early detection of NAFLD is of paramount importance. Some researchers are attempting to explore the correlations between NAFLD and patients' physiological information while designing low-cost detection methods \cite{b6,b7,b8,b9,b10,b11}. Indeed, these methods often require invasive diagnostic procedures for patients \cite{b8,b9,b10,b11} or lack sufficient experimental data on human subjects \cite{b6,b7}. This adds complexity to the diagnosis of NAFLD.

Traditional Chinese Medicine(TCM) tongue diagnosis, with its non-invasive diagnostic approach and rich clinical experience, has gradually gained recognition worldwide. However, its strong subjectivity and ambiguity limit its development. With the advancement of computer technology, intelligent tongue diagnosis techniques can effectively enhance the diagnostic efficiency of TCM tongue diagnosis and provide greater objectivity. Some researchers have indicated correlations between tongue features and certain diseases. They have achieved promising results by using tongue features as diagnostic criteria for diseases and syndromes\cite{b12,b13,b14,b15,b16,b17,b18,b19,b20,b21}. Especially concerning chronic diseases, patients' tongue features exhibit more pronounced characteristics\cite{b11, b13,b15,b16,b17,b20,b21}.

On the basis, the study devised a low-cost and non-invasive NAFLD detection method by integrating the tongue features with easily accessible physiological features. In terms of the fusion approach, this paper adopts an innovative model structure named SelectorNet, which is capable of learning the selection of crucial features and provides interpretive features for model attention. Compared to other detection methods, our method offers lower detection costs and enhanced interpretability, making it convenient for users to perform self-assessment in their daily lives. 

\section{Related Work}

To address the complexity of conventional NAFLD diagnostic methods, some researchers have designed alternative methods for diagnosing NAFLD. \cite{b6} developed individualized models for males and females, employing customized data processing techniques, and utilizing liver function and physiological parameters for the screening of NAFLD. \cite{b7}employed a novel non-invasive tool, photothermal strain imaging (pTSI), for the diagnosis of NAFLD. In the experiment, clinical ultrasound B-mode images were used, and pTSI was used to non-invasively monitor the fat accumulation in the liver of live rats. \cite{b10} utilized multiple sets of effective factors, such as blood flow velocity, psychological, and laboratory test data, and applied machine learning (ML) methods to classify samples into healthy and NAFLD patient groups. \cite{b11} employed a combination of quantitative tongue image features, basic information, and serological indexes, and utilized multiple machine learning methods for NAFLD diagnosis. The experimental results indicate that the application of computer intelligent tongue diagnosis technology can improve the accuracy of NAFLD diagnosis. These methods have lower detection costs compared to conventional diagnostic approaches. However, they often require the patients' serum biomarkers, which adds extra diagnosis expenses. The reliance on professional equipment and personnel still exists for patients to gain insight into their NAFLD condition.

In TCM theory, the color, texture, and other features of the tongue contain valuable information about a patient's constitution. Some researchers have explored the correlations between tongue features and specific diseases, using tongue features as a basis for disease diagnosis. \cite{b13} examined the correlation between captured tongue image features and ALT/AST levels over a period of 2.5 years. The experimental results revealed a strong correlation between certain tongue image features and AST or ALT levels, suggesting that these tongue features may serve as early indicators of liver disease. \cite{b15} proposed a machine learning-based method for diabetes tongue recognition. The tongue images were transformed to the HSV color space, and converted into an image with higher contrast. Thresholding and morphological operations were then applied to the images to segment the tongue area. Finally, the method extracted tongue color features and utilized SVM and KNN for diabetes diagnosis based on image recognition. \cite{b16} developed a computer-aided intelligent decision support model using DenseNet to identify essential features in tongue images. The classification was performed using SVM, and Particle Swarm Optimization (PSO) was used to fine-tune SVM parameters. This method outperforms previous approaches and can be used for early identification of diabetes. \cite{b17}employed a combination of deep neural networks(DNN), support vector machine (SVM), and convolutional neural network (CNN) based on tongue surface and color features to enhance the accuracy of gastric cancer diagnosis. The experimental results demonstrated that the DenseNet architecture outperformed other architectures, achieving a diagnostic accuracy of 91\% in gastric cancer diagnosis. Various experimental results indicate that tongue features have a certain correlation with certain chronic diseases and compounds within the human body. The non-invasive nature of tongue diagnosis also greatly enhances its convenience, implying that tongue diagnosis can be utilized as an early detection method for chronic diseases.

\begin{figure*}[!ht]
\centering
\includegraphics[width=110mm]{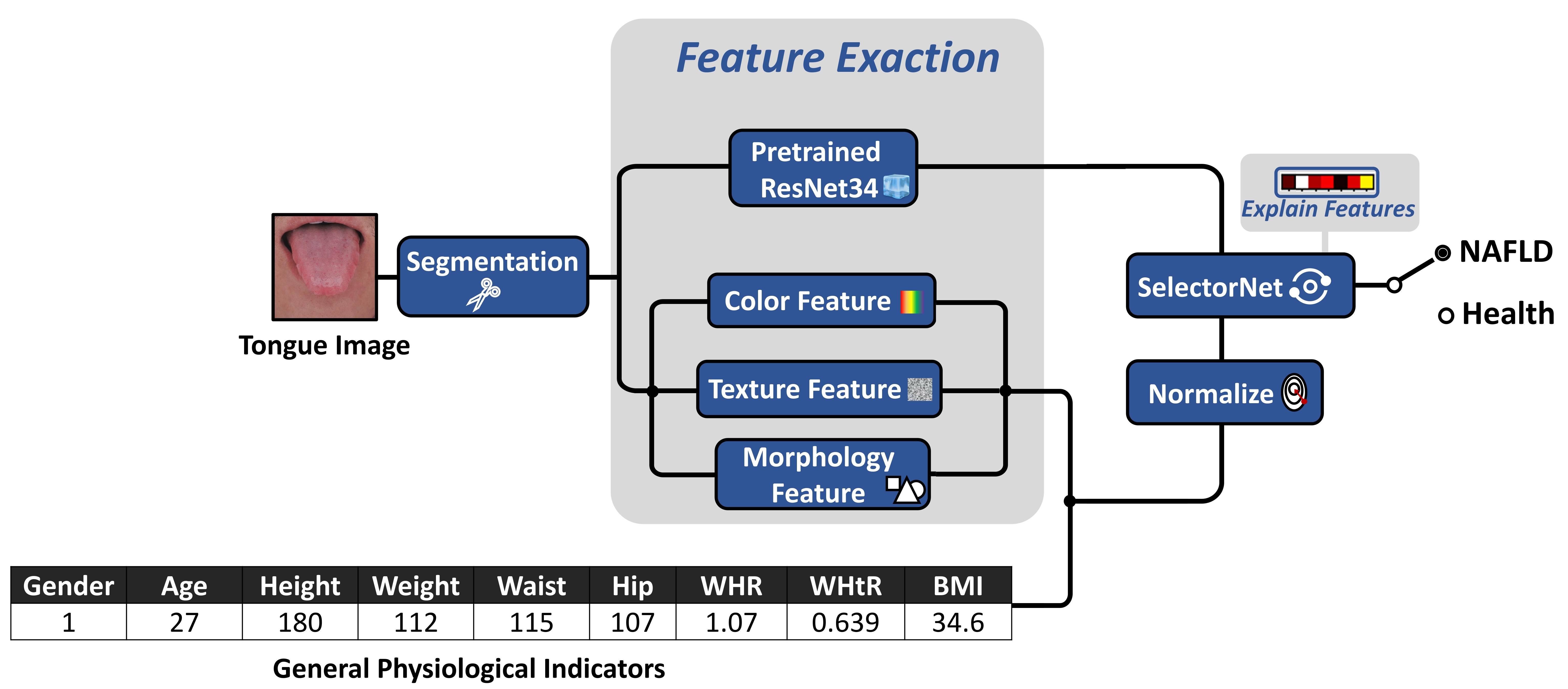}
\caption{The overall framework of the method in this paper.}
\end{figure*}

\section{Method}

The proposed method's overall architecture is illustrated in the Fig. 1. This method requires only the tongue image along with readily obtainable physiological indicators to diagnose NAFLD. Apart from the calculable indicators, the required user-provided indicators are only Gender, Age, Height, Weight, Waist Circumference, and Hip Circumference. Initially, the patient's tongue image is automatically segmented. Subsequently, the segmented tongue image undergoes two modes of feature extraction: one involves custom-defined disease relevance feature extraction methods based on prior research experience, while the other employs a ResNet34 pre-trained on ImageNet21k to generate features, thereby preventing the loss of latent disease-related features. Finally, the custom-extracted tongue image features are concatenated with the patient's physiological indicators as input to SelectorNet. Meanwhile, the features derived from ResNet34 are fused in the final stage of SelectorNet to yield the ultimate diagnostic result. The remaining of this chapter will provide a comprehensive exposition of the methods within this study.

\subsection{Tongue Segmentation}

The diagnostic method proposed in this study possesses low-cost detection capability, making it suitable for patient diagnosing in daily life scenarios. Under such circumstances, a robust tongue segmentation method is required to accommodate complex real-life backgrounds. TongueSAM is a tongue segmentation approach based on the Segment Anything Model (SAM)\cite{b22}, employing YOLOX as the Prompt Generator to reduce the cost of manual prompt. Compared to other methods, TongueSAM demonstrates superior segmentation performance under zero-shot and challenging background conditions, aligning with the application scenario of this study. Furthermore, since TongueSAM employs the same Image Encoder and Prompt Encoder as SAM, and has undergone fine-tuning of the Prompt Generator and Mask Decoder using a substantial amount of tongue image data, it can be directly used without any additional training.

\subsection{Feature Exaction}

According to principles from TCM, tongue features such as color, texture, and morphology contain comprehensive constitutional information. Additionally, research indicates a close correlation between tongue features, particularly tongue color features, and various diseases \cite{b13,b15,b17,b21}. It's noteworthy that despite the existence of numerous automated tongue feature extraction systems\cite{b28,b29,b30,b31}, these methods often require specialized hardware support, introducing additional costs to the overall implementation of the method. Consequently, taking inspiration from the Tongue Diagnosis Analysis System (TDAS) independently developed by Shanghai University of Traditional Chinese Medicine \cite{b31} and commonly considered tongue features in TCM tongue diagnosis, this study designs a set of device-agnostic feature extraction methods, as shown in Fig. 2. This method enhances the flexibility and extensibility of the tongue feature extraction, making it adaptable for use in other diagnostic methods. Furthermore, to prevent the omission of pivotal tongue features, this study incorporates features generated by a ResNet34 pre-trained on ImageNet21k in its final layer. These features are fused in the final stage of SelectorNet, aiming to achieve enhanced diagnostic outcomes.

{
\setlength{\parindent}{0cm}
\textbf{1.Tongue color features}
}

Tongue color, as a widely employed feature, has been utilized for the diagnosis of various diseases and constitutions. Within TCM theory, the color of both the tongue coat and the tongue body carries distinct meanings. Consequently, it's essential to separate and analyze the tongue coat and tongue body respectively. From the perspective of tongue image, the primary differentiation between them lies in their distinct colors, tongue coat usually appears pale white, whereas the tongue body tends to exhibit a light red hue. Therefore, this study segment the tongue coat and tongue body by computing the color differences of each pixel in the RGB space, the calculation formula is shown as follows:

\begin{small}
\begin{equation}
\text{Pixel}
\begin{cases}
    \begin{aligned}
        &\text{Tongue Coat} & ,\text{if } R - (G + B) \leq -45 \\
        &\text{Tongue Body} & ,\text{else}
    \end{aligned}
\end{cases}
\end{equation}
\end{small}

After obtaining separate tongue coat and tongue body, this study calculates their respective mean values of color. Apart from the commonly used RGB color space, the mapping of tongue color to other color spaces has also found widespread use in diagnosing various diseases. Thus, this study projects the colors of the tongue coat and tongue body into three additional color spaces: HIS, YCrCb, and Lab. Beyond color information, the proportion of the tongue coat's holds insights into the patient's health status. A relatively significant proportion of tongue coat coverage could imply excessive dampness or poor digestion within the patient. Hence, this study also extracts the proportion of the tongue coat relative to the entire tongue surface as one of the indicators for NAFLD diagnosis.

\begin{figure}[!ht]
\centering
\includegraphics[width=55mm]{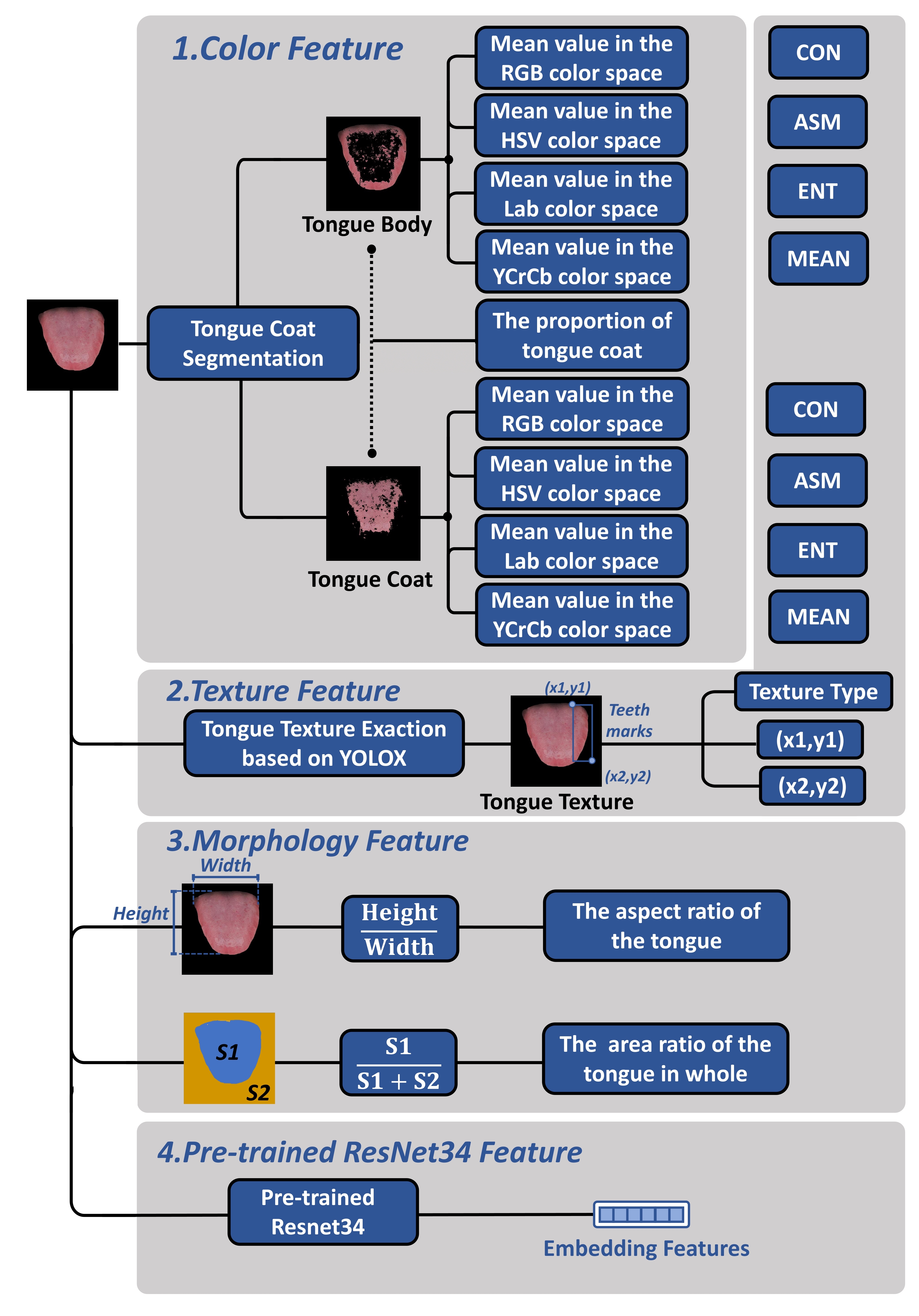}
\caption{The tongue feature extraction method in this paper.}
\end{figure}

{
\setlength{\parindent}{0cm}
\textbf{2.Tongue texture features}
}

Tongue texture features are often comprehensively analyzed in conjunction with tongue color in TCM theory. The significance of tongue texture features in disease diagnosis has garnered gradual attention from various researches\cite{b24,b25}. Inspired by TDAS \cite{b31}, this study employs contrast (CON), angular second moment (ASM), entropy (ENT), and mean (MEAN) to characterize the delicacy of both the tongue body and tongue coat. ASM is inversely related to the other three features. As ASM increases, CON, ENT, and MEAN decrease, indicating reduced delicacy of the tongue body and tongue coat. Similar to the extraction of tongue color, the texture information of tongue images is processed separately for the tongue coat and tongue body.

Additionally, this study also considers tongue texture features that are frequently used in TCM, such as spots, cracks, etc. Texture segmentation poses higher difficulty and annotation costs compared to tongue segmentation. Some texture features are challenging to distinguish with explicit boundaries, such as spots. Thus, this study employs object detection methods to extract tongue texture features, representing the type of texture feature and the coordinates of the detection box, as shown in Fig. 2. To compare the performance of different object detection methods in texture extraction, this study conducts experiments using several commonly used object detection methods. A dataset of 319 tongue images annotated by TCM clinical expert, including cracks, peeled coat, spots, and teeth marks, is selected for experimentation. The dataset is augmented by varying brightness, resulting in a dataset size of 638 images, with an 8:1:1 ratio for training, testing, and validation sets,. The PyTorch framework is used for training, employing consistent training parameters for each model: initial learning rate of $1*10^{-5}$, 30 training epochs, and batch size of 8. The experimental results are presented in the TABLE I.

\begin{table}[htb]   
\begin{center}   
\caption{Comparison of the different tongue texture feature extraction methods.}  
\begin{tabular}{cccc}
\hline   &mIoU & mPA & Acc \\   
\hline   Yolov3& 59.90\% & 68.53\%&67.24\%  \\ 
   Yolov4& 63.55\% & 63.23\%&69.06\%  \\      
   Yolov5& 61.52\% & 76.13\%& 67.47\%  \\    
  FasterRCNN&57.89\% &53.12\%&50.58\%  \\ 
  Yolox& \textbf{82.46\%} &\textbf{ 85.32\%}&\textbf{85.81\%}  \\     
\hline   
\end{tabular}   
\end{center}   
\end{table}

The results demonstrate YOLOX's superiority in the task of texture extraction. This underscores YOLOX's effectiveness in extracting texture information, leading this study to adopt the trained YOLOX as the tool for texture feature extraction within the method.

{
\setlength{\parindent}{0cm}
\textbf{3.Tongue morphology features}
}

In addition to color and texture features, the morphology of the tongue also plays a significant role in TCM tongue diagnosis. Although the acquisition of tongue images through computers limits the observation of tongue motion states, it is possible to characterize the morphology of a patient's tongue based on the proportion of the tongue's area within the overall image and the ratio between the length and width of the tongue in the image. These features can characterize aspects such as the thickness of the tongue, holding valuable insights for disease diagnosis. Therefore, this study incorporates them as part of the morphology feature set extracted from tongue images.

{
\setlength{\parindent}{0cm}
\textbf{4.Tongue features extracted by pre-trained ResNet34}
}

In addition to the commonly used tongue features in diagnosis mentioned above, to ensure that no relevant tongue features related to disease are overlooked, this study employs a ResNet34 pre-trained on ImageNet21K as an additional feature extractor. In this study, each tongue image is processed through the pre-trained ResNet34, extracting features before the final classification layer, which are then dimensionally reduced to 10 dimensions. It's important to note that retaining an excessive number of features can impact the network's ability to focus on more critical features. These features are subsequently merged in the final classification stage of the SelectorNet. 

\begin{figure*}[!ht]
\centering
\begin{adjustbox}{center}
\includegraphics[width=145mm]{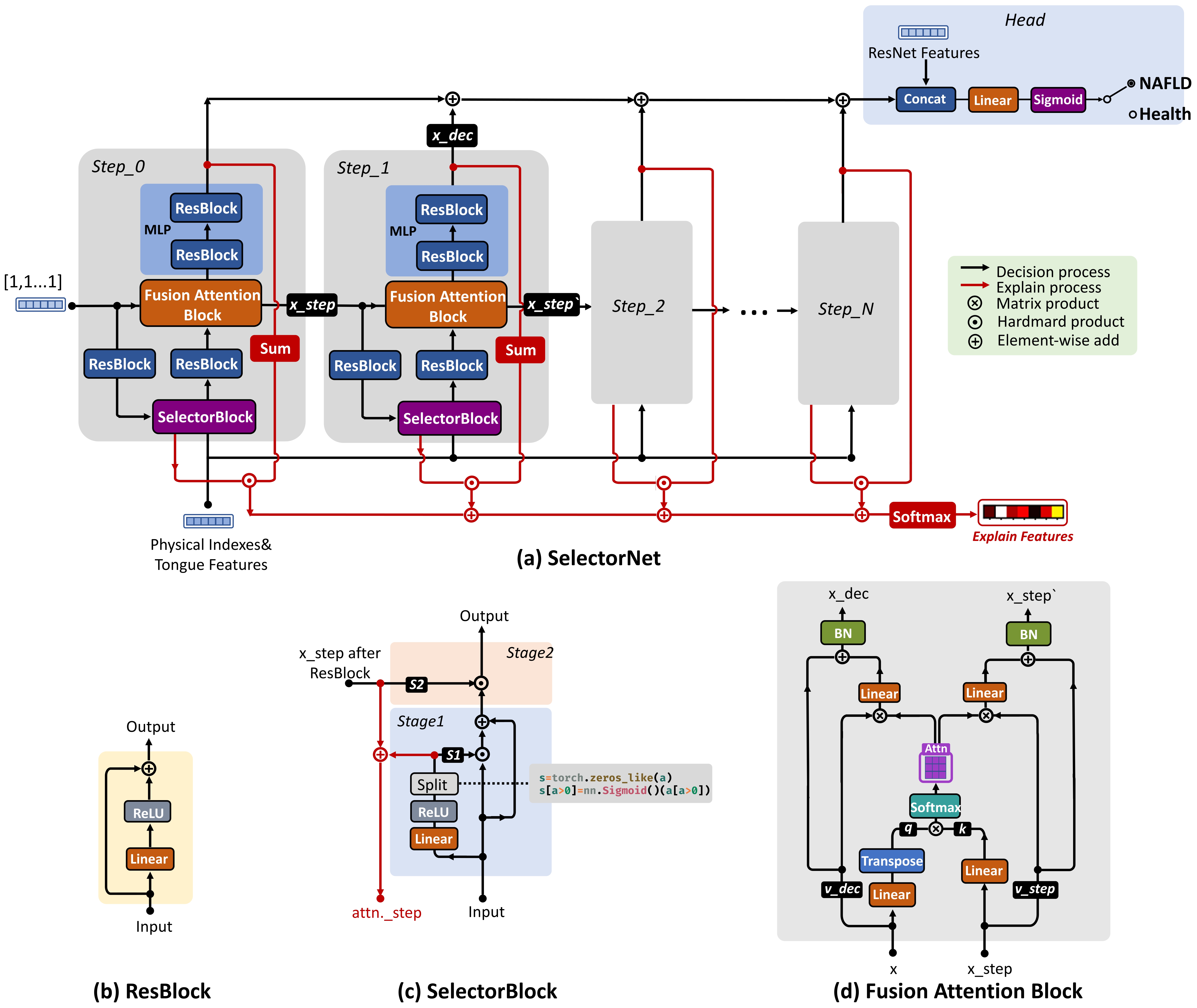}
\end{adjustbox}
\caption{The model structure of SelectorNet. (a) SelectorNet (b) ResBlock (c) SelectorBlock (d) Fusion Atteniton Block}
\end{figure*}

\subsection{SelectorNet}

After the aforementioned feature extraction process, this study concatenates the tongue features with patients' physiological indicators and utilizes them as inputs for the SelectorNet. It is important to note that due to the significant differences among input features, the data is normalized before input the model to mitigate the impact of data with varying scales on the results. The normalized method is computed as follows:

\begin{small}
\begin{equation}
k'=\frac{k-k_{min}}{k_{max}-k_{min}}
\end{equation}
\end{small}

Among them, $k_{max}$ and $k_{min}$ represent the maximum and minimum values of each data category in all data sets respectively, $k$ represent the original data, and $k'$ represent the data after normalization. Through the data normalization, we can map all data to the interval of [0,1]. The structure of SelectorNet, depicted in Fig. 3, encompasses multiple steps inspired by the concepts in TabNet\cite{b26}. The benefits of it can avoid over-parameterization of the model and establish connections between steps, yielding superior performance compared to simple module stacking and multi-headed architectures. With the exception of the initial step, each subsequent step receives two sets of inputs: the original input data and the  x\_step propagated from the previous step. Two sets of outputs will be generated in each step, namely x\_step' for transmission to the next step, and x\_dec for decision-making output. The x\_step is instrumental in aiding the model's selection of diverse features. In the initial step, given the absence of x\_step passed from the previous step, a matrix entirely populated with ones is employed as the initally x\_step.

In each step, the input data is initially passed to feature selection through the SelectorBlock utilizing the processed x\_step from the ResBlock. The SelectorBlock can be partitioned into two stages, both fundamentally constituting feature selection processes, herein named as Stage1 and Stage2. In Stage1, the primary input features undergo transformation via a linear layer and ReLU activations, followed by a Split operation, culminating generate a set of selection matrices denoted as S1. The computational formula for S1 is expressed as follows:

\begin{small}
\begin{equation}
S1=Split(ReLU(Linear(Input))
\end{equation}
\end{small}

The Stage1 selection process is finalized by taking the Hardmard product and the sum of S1 with the input features. The Split operation applying a Sigmoid operation to features greater than 0. The use of Sigmoid is intended to limit the enhancement of features, preventing the generation of extreme values due to excessive amplification. In the case of features equal to 0, the presence of the shortcut structure ensures the retention of the original input features, thus preventing the loss of input features. These features might play a significant role in future computations. 

In Stage2, the features after Stage1 are Hardmard multiplied with the processed x\_step from the ResBlock, and this result is taken as the final output of the SelectorBlock. Therefore, the selection matrix S2 in Stage2 is:  

\begin{small}
\begin{equation}
S2=ResBlock(x_{step})
\end{equation}
\end{small}

This design also draws inspiration from the TabNet, wherein at each step, the selection of features is guided by the information propagated from the preceding step. Within the SelectorBlock, the Stage1 and Stage2 respectively employ the original input data and the information transmitted from the previous step as the basis for feature selection. This enables the model to comprehensively select attributes of input features.

After processing via a ResBlock, the processed features x and the x\_step are fed into the Fusion Attention Block (FAB). The primary function of this module is to fuse the information provided by x\_step with the features of the current step. It generates two sets of features, namely x\_step' for passing to the next step and x\_dec for decision output. The design concept of this module draws inspiration from TabTransformer\cite{b32}. Specifically, we employ linear transformations on x and x\_step to generate the query (q) and key (k) for the attention mechanism. To ensure dimension compatibility, x requires a transposition operation following the linear transformation. The q and k are then multiplied to create an attention matrix reflecting the feature correlations between them. Distinct from conventional attention mechanisms, FAB treats x and x\_step as two types of values, each being multiplied by the same attention matrix. The results undergo feature computation via a linear operation. The final output is connected to the original input through a shortcut, with Batch Normalization ensuring output stability. The generated x\_step' serves as input for the next step, while x\_dec continues through ResBlock to acquire the ultimate output features. Given the differing optimization objectives of x\_dec and x\_step, with x\_dec aiming to enhance decision-making capacity and x\_step to transfer inter-step information, this fusion method promotes a balanced interaction between x\_dec and x\_step. This approach mitigates interference between the two sets of features.

The decision features generated by FAB are further processed through an MLP consisting of two stacked ResBlock to obtain the x\_dec used for decision-making in this step. The x\_dec generated at each step is cumulatively added, and concatenated with the dimension-reduced features generated by ResNet in the Head module. This concatenated features are passed through linear layers and a Sigmoid activation to obtain the final diagnostic result. 

In addition to providing diagnostic outcomes, the proposed SelectorNet in this paper also offers insights into the model's attention to different features. Within each step, attention is computed by multiplying the selection to features and the contribution of that step to the final decision. The selection of features is obtained by adding the two selection matrices, S1 and S2, respectively. The contribution to the results is obtained by summing the features in x\_dec at this step. The computation formula of attention in each step are as follows:

\begin{small}
\begin{equation}
selec.=S1+S2
\end{equation}
\end{small}

\begin{small}
\begin{equation}
contri.=Sum(x_{dec},dim=-1)
\end{equation}
\end{small}

\begin{small}
\begin{equation}
attn._{step}=selec.\cdot contri.
\end{equation}
\end{small}

The overall attention of the model is a sum of the attention from each step. The formula is as follows:

\begin{small}
\begin{equation}
attn._{all}=\sum^{N} attn._{step}\
\end{equation}
\end{small}

The overall attention of the model, presented as visualized features, helps users understand the emphasized attributes and contributes to explaining the model's decision-making process.

\section{Experiments}

From July 2018 to December 2018, this study recruited volunteers from the medical examination center of Shuguang Hospital. A total of 2115 subjects participated in the study, after excluding samples with incomplete clinical data, a total of 1778 cases were included in the research. Among them, there were 831 cases in the NAFLD group and 947 cases in the non-NAFLD group. The inclusion criteria for NAFLD patients were as follows: aged between 25 and 80 years, diagnosed with fatty liver by ultrasound, and not receiving treatment. The following populations were excluded: patients with malignant tumors, metabolic liver diseases, or autoimmune liver diseases; patients who consumed more than 140 grams of alcohol per week for females or more than 210 grams for males in the past 12 months; patients with NAFLD caused by known extrahepatic diseases and/or drug factors; pregnant or lactating women. Additionally, individuals who were not diagnosed with NAFLD, had no history of liver disease, and had normal liver function were included as the non-NAFLD group for analysis. Participants' tongue images were collected using the TFDA-1 tongue diagnostic instrument\cite{b33}, which consists of a CCD device and a standard D50 light source. Participants' routine physiological indicators were collected by the researchers, we selected indicators including Gender, Age, Height, Weight, Waist Circumference, Hip Circumference, WHR, WHtR, and BMI. It's worth mentioning that the dataset used in this experiment is the same as \cite{b11}, which is the result of collaboration between this study and Shanghai University of Traditional Chinese Medicine.

All experiments in this paper were conducted using the same hardware configuration, with an Intel(R) Core(TM) i9-7940X CPU, 16GB of memory, and a GeForce RTX 2080 Ti graphics card. The training framework chosen is PyTorch, and all models were validated using K-fold cross-validation (K=5). Patient's tongue image features were extracted using the method presented in this paper, and these features were concatenated with the patient's routine physiological indicators to form the input for the models. The evaluation metrics used for assessing include Accuracy, Precision, Recall, and Specificity. The rest of hyperparameter settings for different experiments will be introduced in respective sections.

\subsection{Compared with other classify methods.}

To validate the performance of the diagnostic method proposed in this paper, several advanced structured data classification models were selected and compared with the SelectorNet\cite{b34}. For a fair evaluation, we employed the Optuna library\cite{b35} to perform hyperparameter tuning for each model over 100 iterations, early stopping technique was employed to prevent overfitting. For SelectorNet, the number of steps was 3, the initial learning rate was set at 0.4637, the training epochs were set to 584, and the learning rate was converged using the cosine annealing method. The hyperparameters for the other methods are available in our github project. The experimental results are presented in the TABLE II.

\begin{table*}[htb]   
\begin{center}   
\caption{Comparison of the different classify methods.}  
\begin{tabular}{ccccc}
\hline   &Accuracy & Precision & Recall & Specificity \\   
\hline   
LinearModel&75.03\% $\pm$ 2.22\% & 74.97\% $\pm$ 1.34\% & 79.62\% $\pm$ 0.69\% & 69.79\% $\pm$ 4.84\% \\
KNN&66.87\% $\pm$ 1.15\% & 67.90\% $\pm$ 1.03\% & 77.09\% $\pm$ 1.36\% & 55.23\% $\pm$ 3.01\% \\
SVM&70.81\% $\pm$ 1.60\% & 67.82\% $\pm$ 2.50\% & 69.91\% $\pm$ 3.90\% & 71.84\% $\pm$ 2.92\% \\
DecisionTree&66.70\% $\pm$ 1.62\% & 65.02\% $\pm$ 2.50\% & 70.22\% $\pm$ 4.07\% & 62.69\% $\pm$ 3.45\% \\
RandomForest&74.02\% $\pm$ 1.83\% & 74.78\% $\pm$ 1.09\% & 80.25\% $\pm$ 0.45\% & 66.90\% $\pm$ 4.12\% \\
XGBoost&74.50\% $\pm$ 3.30\% & 74.45\% $\pm$ 3.30\% & 78.67\% $\pm$ 2.72\% & 71.12\% $\pm$ 4.55\% \\
CatBoost&74.80\% $\pm$ 2.19\% & 74.25\% $\pm$ 2.05\% & 78.56\% $\pm$ 1.65\% & 70.51\% $\pm$ 3.72\% \\
LightGBM&75.09\% $\pm$ 2.98\% & 74.56\% $\pm$ 3.29\% & 78.78\% $\pm$ 2.90\% & 70.88\% $\pm$ 4.07\% \\
ModelTree&75.25\% $\pm$ 1.79\% & 75.06\% $\pm$ 1.25\% & 79.52\% $\pm$ 0.73\% & 70.39\% $\pm$ 3.76\% \\
MLP&75.14\% $\pm$ 1.67\% & 76.20\% $\pm$ 1.77\% & 81.31\% $\pm$ 1.85\% & 68.11\% $\pm$ 3.74\% \\
TabNet&74.69\% $\pm$ 2.26\% & 71.98\% $\pm$ 1.34\% & 74.45\% $\pm$ 1.17\% & \textbf{74.97\% $\pm$ 5.15\%} \\
VIME&73.34\% $\pm$ 2.57\% & 71.78\% $\pm$ 2.08\% & 75.61\% $\pm$ 2.27\% & 70.76\% $\pm$ 5.07\% \\
TabTransformer&74.69\% $\pm$ 1.87\% & 74.74\% $\pm$ 1.94\% & 79.41\% $\pm$ 2.35\% & 69.31\% $\pm$ 4.61\% \\
RLN&66.29\% $\pm$ 2.73\% & 60.46\% $\pm$ 3.51\% & 74.06\% $\pm$ 2.65\% & 54.62\% $\pm$ 2.82\% \\
STG&74.97\% $\pm$ 2.32\% & 73.25\% $\pm$ 2.34\% & 76.45\% $\pm$ 3.22\% & 73.28\% $\pm$ 5.18\% \\
NAM&74.52\% $\pm$ 1.97\% & 74.39\% $\pm$ 1.88\% & 79.09\% $\pm$ 1.38\% & 69.31\% $\pm$ 3.26\% \\
DeepFM&75.36\% $\pm$ 2.13\% & 75.00\% $\pm$ 1.82\% & 79.30\% $\pm$ 1.37\% & 70.88\% $\pm$ 4.01\% \\
SAINT&74.97\% $\pm$ 2.05\% & 73.73\% $\pm$ 1.39\% & 77.51\% $\pm$ 1.27\% & 72.08\% $\pm$ 4.42\% \\
\hline   
SelectorNet&\textbf{77.22\% $\pm$ 1.56\%} & \textbf{78.81\% $\pm$ 2.50\% }& \textbf{83.32\% $\pm$ 2.64\% }& 70.28\% $\pm$ 3.55\% \\
SelectorNet\_only Tongue Feature&61.81\% $\pm$ 1.84\% & 60.09\% $\pm$ 2.72\% & 67.06\% $\pm$ 3.96\% & 55.82\% $\pm$ 2.49\% \\
SelectorNet\_only Physiological Indicators&75.48\% $\pm$ 1.93\% & 74.46\% $\pm$ 1.61\% & 78.24\% $\pm$ 1.60\% & 72.32\% $\pm$ 4.17\% \\
\hline
\end{tabular}   
\end{center}   
\end{table*}

The experimental results demonstrate that SelectorNet exhibits a leading advantage across most evaluation metrics. In terms of the Accuracy evaluation metric, SelectorNet approximately achieves a 2\% improvement compared to other methods, showcasing its superiority. The observations considering only tongue features and only physiological indicators reflect the significance of incorporating tongue features into diagnosis and SelectorNet's ability to fuse these two types of features. Tongue features contain valuable disease-related information about patients, assisting the model in making accurate diagnoses.

\subsection{Interpretation.}

To validate the interpretability of SelectorNet, we selected the best-performing fold model from the k-fold validation in the previous section for analysis. We randomly selected several samples from the test set and observed SelectorNet's attention to different features. The result is depicted in the Fig. 4.

Based on the interpretability matrix generated by SelectorNet, it is evident that Weight, Waist Circumference, and BMI have become crucial features that the model pays attention to, aligning with our common knowledge. Furthermore, certain tongue features, particularly texture features, have also gained significant attention from the model. This underscores the importance of tongue features as diagnostic indicators.

To further test the model's interpretability, we introduced an additional 30 randomly distributed perturbation attributes to the original inputs and examined how the model's attention changed. The result is illustrated in the Fig. 5.

The experimental results indicate that even in the presence of noise, SelectorNet is capable of concentrating its primary attention on more important features, such as Weight, Waist and BMI. This illustrates the persuasive interpretability and notable selection ability by SelectorNet.

\begin{figure}[!ht]
\centering
\includegraphics[width=60mm]{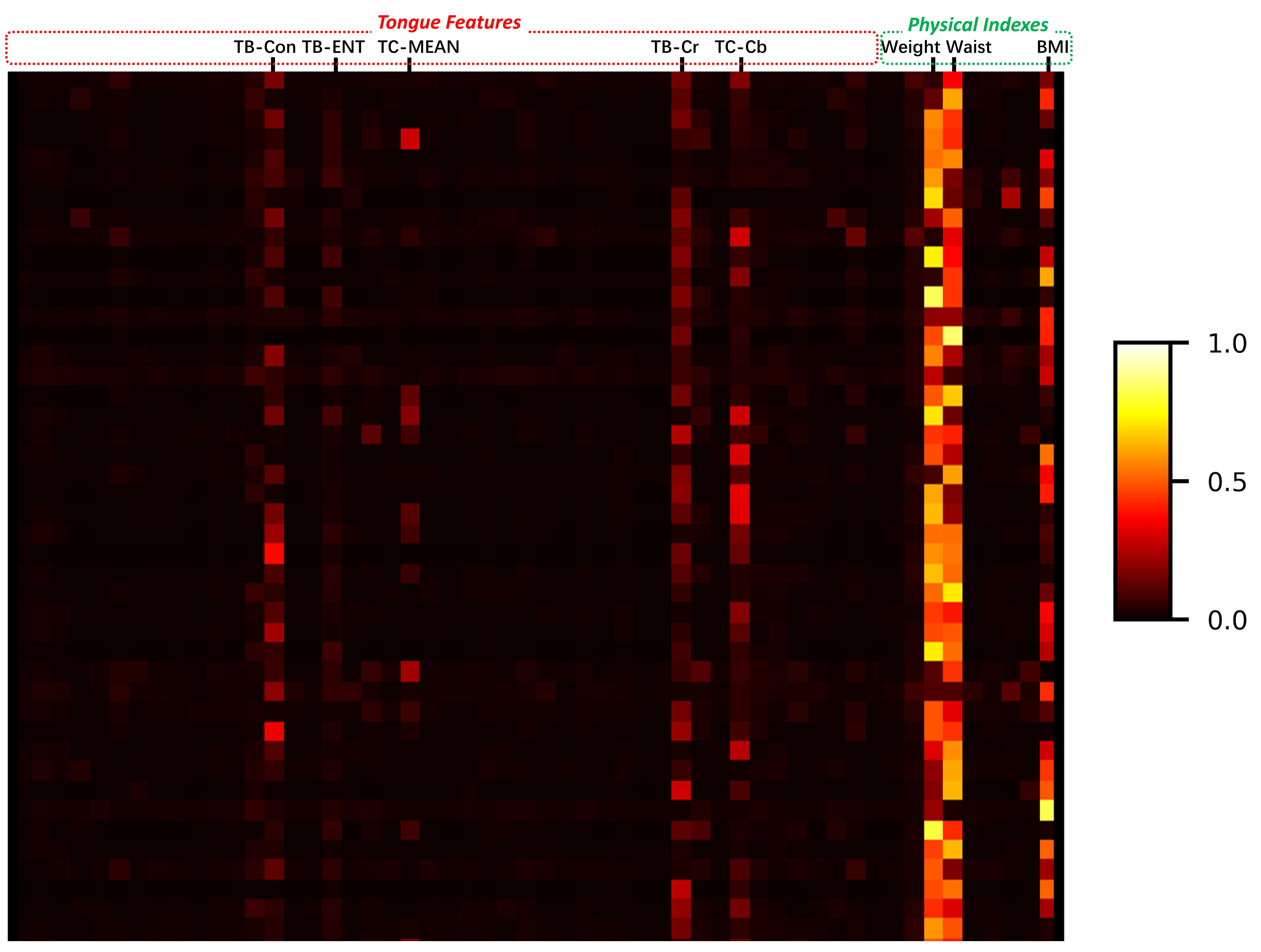}
\caption{The interpretive matrix generated by the model for the test samples.}
\end{figure}

\begin{figure}[!ht]
\centering
\includegraphics[width=80mm]{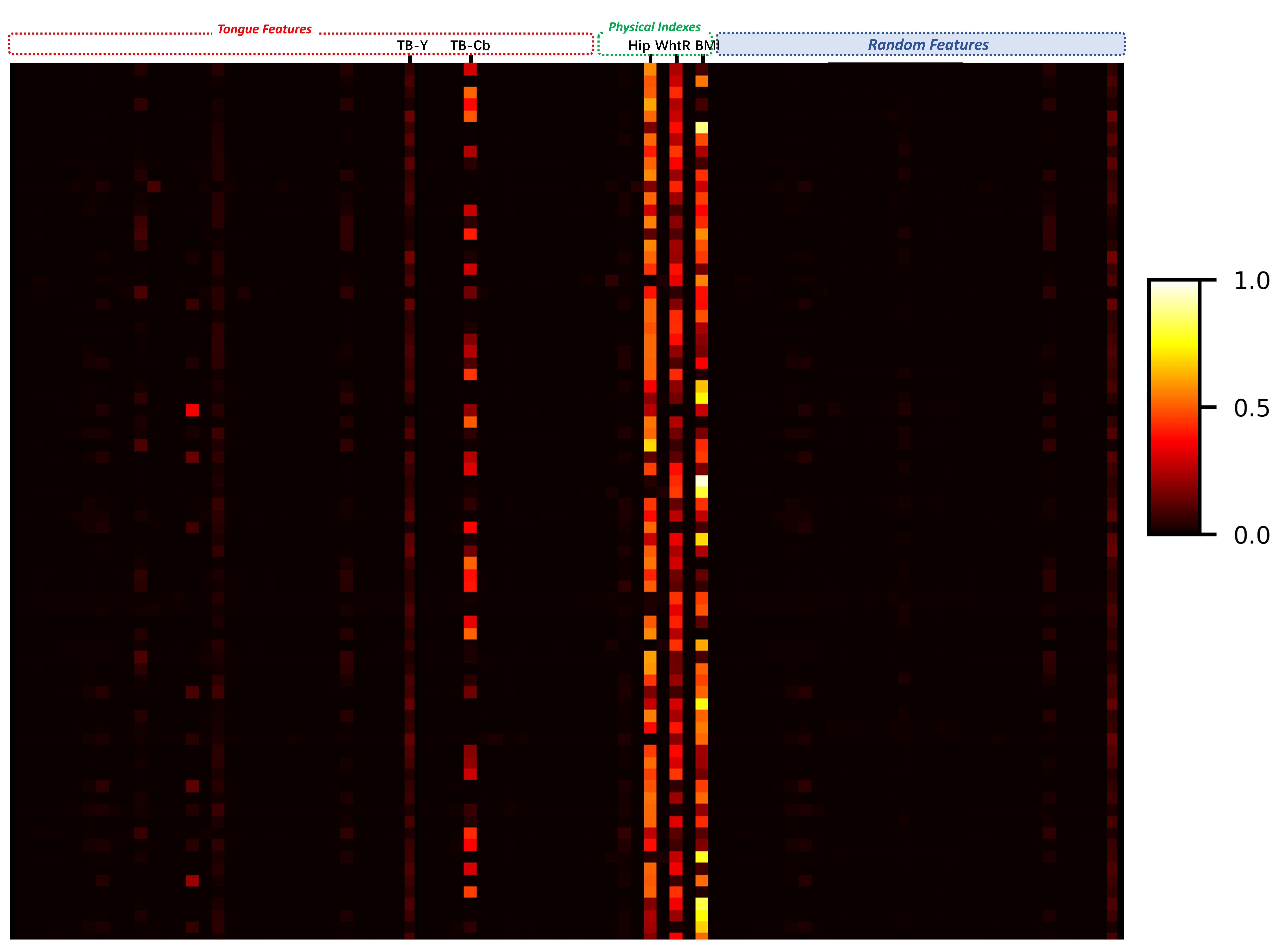}
\caption{The interpretive matrix fused with noise generated by the model for the test samples.}
\end{figure}

\subsection{Ablation experiment.}

To validate the effectiveness of the proposed modules in this paper, ablation experiments were conducted on SelectorNet. For the ResBlock, SelectorBlock, and FAB modules, common alternatives were tested for each. The experimental parameters were set the same as in the first section, and the results are presented in the TABLE III.

\begin{table*}[htb]   
\begin{center}   
\caption{Ablation experiment.}  
\begin{tabular}{cccccc}
\hline   origin module&replaced by&Accuracy & Precision & Recall & Specificity \\   
\hline   
ResBlock&Linear+ReLU&76.71\% $\pm$ 1.91\% & 76.77\% $\pm$ 2.70\% & 80.78\% $\pm$ 2.75\% & 72.08\% $\pm$ 2.14\% \\
\hline   
&Concat+Linear+ReLU&76.04\% $\pm$ 2.76\% & 74.73\% $\pm$ 3.05\% & 78.04\% $\pm$ 3.59\% & \textbf{73.77\% $\pm$ 5.20\%} \\
&Add+Linear+ReLU&76.24\% $\pm$ 2.02\% & 77.29\% $\pm$ 2.03\% & 81.52\% $\pm$ 1.65\% & 71.72\% $\pm$ 3.39\% \\
SelectorBlock&HardMard+Linear+ReLU&76.60\% $\pm$ 1.98\% & 76.50\% $\pm$ 2.75\% & 80.46\% $\pm$ 2.82\% & 72.20\% $\pm$ 2.59\% \\
&only Stage1&75.53\% $\pm$ 1.66\% & 74.61\% $\pm$ 2.23\% & 78.35\% $\pm$ 2.44\% & 72.33\% $\pm$ 1.78\% \\
&only Stage2&76.40\% $\pm$ 2.21\% & 77.03\% $\pm$ 2.97\% & 80.99\% $\pm$ 3.02\% & 72.32\% $\pm$ 3.92\% \\
\hline   
Fusion Attention Block&Concat+Linear+ReLU&71.65\% $\pm$ 3.04\% & 74.96\% $\pm$ 6.87\% & 80.02\% $\pm$ 4.03\% & 62.11\% $\pm$ 2.42\% \\
\hline
No Change&No Change&\textbf{77.22\% $\pm$ 1.56\%} & \textbf{78.81\% $\pm$ 2.50\% }& \textbf{83.32\% $\pm$ 2.64\% }& 70.28\% $\pm$ 3.55\% \\
\hline
\end{tabular}   
\end{center}   
\end{table*}

The experimental results demonstrate that the modules designed in this paper exhibit higher effectiveness compared to other commonly replaceable structures. The central concept of this paper is to minimize feature loss to the greatest extent possible, while also integrating feature selection to provide heightened attention to significant features. This design can result in better performance compared to stacking linear layers.

\section{Conclusion}

This paper introduces a cost-effective and interpretable diagnostic method for NAFLD, integrating TCM tongue features. This method requires only easily accessible physiological indicators and tongue image from patients, enabling accurate diagnosis of NAFLD. The fusion model employed in this study named SelectorNet, can autonomously learn feature selection and avoiding feature loss, leading to superior performance. The model's attention on different attributes can also be visualized.In the future, we plan to expand its applications for user-friendly convenience.

\section{Acknowledge}
This work was supported by Industry-University-Research Cooperation Project of Fujian Science and Technology Planning (No:2022H6012),Natural Science Foundation of Fujian Province of China (No.2021J011169, No.2022J011224), Science and Technology Research Project of Jiangxi Provincial Department of Education (No.GJJ2206003). Special thanks to my friends Xinmiao Zhao, Zhiwei He and Quanxing Xu for their encouragement.

\vspace{12pt}

\begin{thebibliography}{00}
\bibitem{b1}Younossi Z M,  Koenig A B,  Abdelatif D, et al. Global epidemiology of nonalcoholic fatty liver disease—Meta‐analytic assessment of prevalence, incidence, and outcomes[J]. Hepatology, 2016, 64(1).
\bibitem{b2}Liver Fibrosis, but No Other Histologic Features, Is Associated With Long-term Outcomes of Patients With Nonalcoholic Fatty Liver Disease[J]. Gastroenterology, 2015, 149(2):389-397.e10.
\bibitem{b3}Increased risk of mortality by fibrosis stage in nonalcoholic fatty liver disease: Systematic review and meta analysis[J]. Hepatology, 2017, 65(5).
\bibitem{b4}Campana L,  Iredale J. Regression of Liver Fibrosis[J]. Seminars in Liver Disease, 2017, 58(01):001-010.
\bibitem{b5}Nina Gyorfi, Ákos Odry, Zoltán Karádi, Péter Odry, Andras Vereczkei, Bojan Kuljic, Zoltan Vizvari, Attila Tóth:Proof of Concept Clinical Trial of Bioimpedance-based NAFLD Diagnosis Technique. SACI 2021: 141-146
\bibitem{b6}Nina Gyorfi, Ákos Odry, Zoltán Karádi, Péter Odry, Andras Vereczkei, Bojan Kuljic, Zoltan Vizvari, Attila Tóth: Proof of Concept Clinical Trial of Bioimpedance-based NAFLD Diagnosis Technique. SACI 2021: 141-146
\bibitem{b7}Changhoon Choi, Wonseok Choi, Jeesu Kim, Chulhong Kim: Non-Invasive Photothermal Strain Imaging of Non-Alcoholic Fatty Liver Disease in Live Animals. IEEE Trans. Medical Imaging 40(9): 2487-2495 (2021)
\bibitem{b8}Suranjan Panigrahi, Ridhi Deo, Edward A. Liechty:A New Machine Learning-Based Complementary Approach for Screening of NAFLD (Hepatic Steatosis). EMBC 2021: 2343-2346
\bibitem{b9}Shiva Shankar Reddy, Nilambar Sethi, R. Rajender, Gadiraju Mahesh: Forecasting Diabetes Correlated Non-alcoholic Fatty Liver Disease by Exploiting Naïve Bayes Tree. EAI Endorsed Trans. Scalable Inf. Syst. 10(1): e2 (2022)
\bibitem{b10}Afrooz Arzehgar, Raheleh Ghouchan Nezhad Noor Nia, Vajiheh Dehdeleh, Fatemeh Roudi, Saeid Eslami: Non-Alcoholic Fatty Liver Disease Diagnosis with Multi-Group Factors. ICIMTH 2023: 503-506
\bibitem{b11}Tao Jiang, Xiao-jing Guo, Liping Tu, Zhou Lu, Ji Cui, Xuxiang Ma, Xiaojuan Hu, Xinghua Yao, Longtao Cui, Yongzhi Li, Jingbin Huang, Jiatuo Xu: Application of computer tongue image analysis technology in the diagnosis of NAFLD. Comput. Biol. Medicine 135: 104622 (2021)
\bibitem{b12}Meng Xiao, Guozheng Liu, Yu Xia, Hao Xu: A Deep Learning Approach for Tongue Diagnosis. AMLTA 2019: 3-12
\bibitem{b13}Min-Chun Hu, Kun-Chan Lan, Wen-Chieh Fang, Yu-Chia Huang, Tsung-Jung Ho, Chun-Pang Lin, Ming-Hsien Yeh, Paweeya Raknim, Ying-Hsiu Lin, Ming-Hsun Cheng, Yi-Ting He, Kuo-Chih Tseng: Automated tongue diagnosis on the smartphone and its applications. Comput. Methods Programs Biomed. 174: 51-64 (2019)
\bibitem{b14}Zibo Zhou, Dongliang Peng, Fumeng Gao, Lu Leng: Medical Diagnosis Algorithm Based on Tongue Image on Mobile Device. J. Multim. Inf. Syst. 6(2): 99-106 (2019)
\bibitem{b15}Xiaohui Lin, Zhaochai Yu, Zuoyong Li, Weina Liu: Machine Learning Based Tongue Image Recognition for Diabetes Diagnosis. ML4CS (3) 2020: 474-484
\bibitem{b16}S. N. Deepa, Abhik Banerjee: Intelligent decision support model using tongue image features for healthcare monitoring of diabetes diagnosis and classification. Netw. Model. Anal. Health Informatics Bioinform. 10(1): 41 (2021)
\bibitem{b17}Elham Gholami, Seyed Reza Kamel Tabbakh, Maryam Kheirabadi:Increasing the accuracy in the diagnosis of stomach cancer based on color and lint features of tongue. Biomed. Signal Process. Control. 69: 102782 (2021)
\bibitem{b18}Ye Yuan, Wei Liao: Design and Implementation of the Traditional Chinese Medicine Constitution System Based on the Diagnosis of Tongue and Consultation. IEEE Access 9: 4266-4278 (2021)
\bibitem{b19}Romany Fouad Mansour, Maha M. Althobaiti, Amal Adnan Ashour: Internet of Things and Synergic Deep Learning Based Biomedical Tongue Color Image Analysis for Disease Diagnosis and Classification. IEEE Access 9: 94769-94779 (2021)
\bibitem{b20}Qingbin Zhuang, Senzhong Gan, Liangyu Zhang: Human-computer interaction based health diagnostics using ResNet34 for tongue image classification. Comput. Methods Programs Biomed. 226: 107096 (2022)
\bibitem{b21}Nannan Zhang, Zhixing Jiang, JinXing Li, David Zhang: Multiple color representation and fusion for diabetes mellitus diagnosis based on back tongue images. Comput. Biol. Medicine 155: 106652 (2023)
\bibitem{b22}Shan Cao, Qunsheng Ruan, and Qingfeng Wu. "TongueSAM: An Universal Tongue Segmentation Model Based on SAM with Zero-Shot." 2023. arXiv preprint arXiv:2308.06444.
\bibitem{b23}Keye Zhang, Xinfeng Zhang, Farooq Ahmad: Tongue Image Texture Classification Based on Xception. ICCPR 2020: 261-266
\bibitem{b24}Hong-Kai Zhang, Yang-Yang Hu, Xue-Li, Li-Juan Wang, Wen-Qiang Zhang, Fu-Feng Li: Computer Identification and Quantification of Fissured Tongue Diagnosis. BIBM 2018: 1953-1958
\bibitem{b25}Jianqiang Peng, Xinlei Li, Dawei Yang, Yingtao Zhang, Wei Zhang, Ye Zhang, Yajie Kong, Fufeng Li, Wenqiang Zhang: Automatic Tongue Crack Extraction For Real-Time Diagnosis. BIBM 2020: 694-699
\bibitem{b26}Sercan Ö. Arik, Tomas Pfister: TabNet: Attentive Interpretable Tabular Learning. AAAI 2021: 6679-6687
\bibitem{b27}Xin Huang, Ashish Khetan, Milan Cvitkovic, Zohar S. Karnin: TabTransformer: Tabular Data Modeling Using Contextual Embeddings. CoRR abs/2012.06678 (2020)
\bibitem{b28}C.C. Chiu, A novel approach based on computerized image analysis for traditional Chinese medical diagnosis of the tongue, Comput. Methods Progr. Biomed. 61 (2) (2000) 77–89.
\bibitem{b29}Y. Wang, et al., Tongue image color recognition in traditional Chinese medicine, Sheng wu yi xue gong cheng xue za zhi = J. Biomed. Eng. = Shengwu yixue gongchengxue zazhi 22 (6) (2005) 1116–1120.
\bibitem{b30}H. Z. Zhang, K. Q. Wang, D. Zhang, B. Pang and B. Huang, "Computer Aided Tongue Diagnosis System," 2005 IEEE Engineering in Medicine and Biology 27th Annual Conference, Shanghai, China, 2005, pp. 6754-6757, doi: 10.1109/IEMBS.2005.1616055.
\bibitem{b31}Jun Li, Jingbin Huang, Tao Jiang, Liping Tu, Longtao Cui, Ji Cui, Xuxiang Ma, Xinghua Yao, Yulin Shi, Sihan Wang, Yu Wang, Jiayi Liu, Yongzhi Li, Changle Zhou, Xiaojuan Hu, Jiatuo Xu: A multi-step approach for tongue image classification in patients with diabetes. Comput. Biol. Medicine 149: 105935 (2022)
\bibitem{b32}Xin Huang, Ashish Khetan, Milan Cvitkovic, Zohar S. Karnin: TabTransformer: Tabular Data Modeling Using Contextual Embeddings. CoRR abs/2012.06678 (2020)
\bibitem{b33}W. Jiao, X.J. Hu, L.P. Tu, et al., Tongue color clustering and visual application based on 2D information, Int J Comput Assist Radiol Surg 15 (2) (2020) 203–212, https://doi.org/10.1007/s11548-019-02076-z.
\bibitem{b34}Vadim Borisov, Tobias Leemann, Kathrin Seßler, Johannes Haug, Martin Pawelczyk, Gjergji Kasneci: Deep Neural Networks and Tabular Data: A Survey. CoRR abs/2110.01889 (2021)
\bibitem{b35}T. Akiba, S. Sano, T. Yanase, T. Ohta, and M. Koyama, “Optuna: A next- generation hyperparameter optimization framework,” in Proceedings of the 25rd ACM SIGKDD International Conference on Knowledge Discovery and Data Mining, 2019.


\end{thebibliography}
\end{document}